\documentclass[%
 reprint,
nofootinbib,
 amsmath,amssymb,
 aps,
prl
]{revtex4-2}

\usepackage{graphicx}
\usepackage{dcolumn}
\usepackage{bm}
\usepackage{xcolor}
\usepackage{hyperref}
\usepackage{amsmath}
\usepackage{slashed}
\usepackage{cleveref}

\begin{document}

\preprint{APS/123-QED}

\title{Explaining the PeV Neutrino Fluxes at KM3NeT and IceCube\\with Quasi-Extremal Primordial Black Holes}

\author{Michael J.~Baker}
\email{mjbaker@umass.edu}
\author{Joaquim Iguaz Juan}
\email{jiguazjuan@umass.edu}
\author{Aidan Symons}
\email{asymons@umass.edu}
\author{Andrea Thamm}
\email{athamm@umass.edu}
\affiliation{Department of Physics, University of Massachusetts, Amherst, MA 01003, USA
}%

\date{\today}

\begin{abstract}

The KM3NeT experiment has recently observed a neutrino with an energy around 100\,PeV, and IceCube has detected five neutrinos with energies above 1\,PeV.  While there are no known astrophysical sources, exploding primordial black holes could have produced these high-energy neutrinos. For Schwarzschild black holes this interpretation results in tensions between the burst rates inferred from the KM3NeT and IceCube observations, with indirect constraints from the extragalactic gamma ray background and with the non-observation of an associated gamma ray signal at LHAASO.  In this letter we show that if there is a population of primordial black holes charged under a new dark $u(1)$ symmetry which spend most of their time in a quasi-extremal state, the neutrino emission at 1\,PeV may be more suppressed than at 100\,PeV. The burst rates implied by the KM3NeT and IceCube observations and the indirect constraints can then all be consistent at $1\sigma$, and no associated gamma-ray signal was expected at LHAASO. Furthermore, these black holes could constitute all of the observed dark matter in the universe. 

\end{abstract}

\maketitle


\section{Introduction}

The KM3NeT Collaboration recently reported the detection of an extremely high-energy neutrino, $E_{\nu}\sim220$\,PeV~\cite{KM3NeT:2025npi}. This event, KM3-230213A, is the highest energy neutrino ever reported.  The IceCube Collaboration has also reported high energy neutrino events, at the PeV scale~\cite{IceCube:2013cdw,IceCube:2014stg,IceCube:2016umi,IceCube:2021rpz} (see also Ref.~\cite{IceCube:2025ezc}). The IceCube observatory has been observing the sky for a longer time and with a larger effective area than KM3NeT, which raises the question: why did KM3NeT observe such a high energy neutrino before IceCube? If the neutrino is assumed to have come from the diffuse isotropic neutrino flux, this leads to a $3.5\sigma$ tension between the two experiments~\cite{Li:2025tqf}. The most favoured scenario is a transient point source, which reduces the tension to $2\sigma$~\cite{Li:2025tqf} and points to a likely new astrophysical source.

In this work we study the possibility that these neutrinos came from exploding primordial black holes (PBHs).  PBHs are black holes that formed in the early universe.  The most widely studied production mechanism is formation from the collapse of large overdensities~\cite{Hawking:1982ga, Carr:1975qj,Niemeyer:1999ak,Ballesteros:2017fsr,Garcia-Bellido:2017mdw,Sasaki:2025frv}, but other formation mechanisms have been explored in recent years~\cite{Deng:2017uwc,Kusenko:2020pcg,Brandenberger:2021zvn,Baker:2021nyl, Baker:2021sno,Ferreira:2024eru,Ballesteros:2024hhq}.  Once formed, these black holes are expected to emit Hawking radiation and lose mass.  Since their temperature is inversely proportional to their mass~\cite{Hawking:1975vcx}, this leads to a runaway explosion~\cite{Hawking:1974rv}.  While the solar mass black holes observed by binary mergers are so massive that they will not explode any time soon, lighter PBHs could be exploding today.  Experiments such as HAWC search for signals of these explosions~\cite{HAWC:2019wla}, and if one were observed it would provide evidence of the existence of PBHs, evidence of Hawking radiation, and provide definitive information on the particles present in nature~\cite{Baker:2021btk,Baker:2022rkn}.

PBHs have been considered as the high energy neutrino source in previous works.  Ref.~\cite{Klipfel:2025jql} studies whether exploding Schwarzschild PBHs could be the source of the high-energy neutrinos.  They find that the PBH burst rate required to explain the KM3NeT event is significantly larger than that implied by the IceCube events.  Both burst rates are also much larger than the maximum rate allowed by indirect constraints from the Extragalactic Gamma Ray Background (EGRB), although some of these rates are consistent at around $2\sigma$ due to large error bars. A further tension of a Schwarzschild PBH as the neutrino source is with the non-observation of an associated gamma-ray signal in LHAASO seven to fourteen hours before the KM3-230213A event \cite{Airoldi:2025opo}.
Refs.~\cite{Boccia:2025hpm} consider PBHs in the context of the `memory-burden' hypothesis, which suggests that back-reaction effects onto the quantum state of the black hole might become relevant as the PBH evaporates.  This leads to a slowing of the evaporation process which significantly extends the PBH lifetime. However, in this scenario, indirect constraints on the PBH abundance lead to a small expected KM3NeT event rate, making KM3-230213A an unlikely observation (however, see also Refs.~\cite{Zantedeschi:2024ram,Dvali:2025ktz}). 

In this letter we discuss an alternative based on the scenario where a new dark sector is introduced and PBHs are formed with a small initial dark charge~\cite{Baker:2025zxm}. This scenario can yield quasi-extremal primordial black holes, which are very light but cosmologically long-lived. This in turn leads to large local burst rates in the vicinity of the Earth which are compatible with indirect constraints on the PBH abundance. We show how this scenario can naturally explain the PBH burst rates implied by both KM3NeT and IceCube, while remaining consistent with indirect bounds on the PBH population. 
Due to the fast transition from quasi-extremality to the final Schwarzschild-like burst, a gamma-ray signal is only expected within hundreds of seconds before the neutrino event, reconciling this scenario with the non-observation of a gamma-ray signal hours before the neutrino event. 
We also find that this population of PBHs could account for 100\% of the dark matter. 

\section{Quasi-Extremal Primordial Black Holes}
\label{Sec:constraints}

Our current understanding of black hole thermodynamics suggests that PBHs have a temperature inversely proportional to their mass, $T_{\rm PBH}\sim 1/M_{\rm PBH}$~\cite{Hawking:1975vcx}. In addition, their temperature also depends on the charge and spin of the black hole. For charged black holes (known as Reissner–Nordstr\"om (RN) black holes), the temperature is given by~\cite{Page:1976df}
\begin{align}
    T_{\rm PBH} = \frac{M_{\rm Pl}^{2}}{2 \pi M_{\rm PBH}} \frac{\sqrt{1 - (Q^{*})^{2}}}{\left( 1 + \sqrt{1 - (Q^{*})^{2}} \right)^{2}},
    \label{eq:Tbh}
\end{align}
where the charge parameter is defined as $Q^* \equiv Q M_{\rm Pl}/M_\text{PBH}$, $Q$ is the charge of the PBH and we are using units where $c=\hbar=4\pi\epsilon_0=1$ so $M_{\rm Pl} = 1.22 \times 10^{19}$\,GeV.
Roughly speaking, PBHs will emit all particles with masses below the black hole temperature with (almost) a blackbody spectrum.  More precisely, the spectrum of emitted neutral particles per unit time is given by
\begin{align}
   \frac{d^{2}N_p^{i}}{dEdt} = \frac{n_{\rm dof}^{i}}{2\pi}\frac{\Gamma^{i}(M_{\rm PBH},Q,E)}{(e^{E/T_{\rm PBH}} - (-1)^{2s})} \,,
    \label{eq:d2NdEdt}
\end{align}
where $\Gamma^{i}(M_{\rm PBH},Q,E)$ is a greybody factor which depends on the mass and charge of the PBH as well as the energy $E$ and spin $s$ of the emitted particle $i$, and where $n_{\rm dof}^{i}$ is the number of degrees of freedom of particle $i$.

In a realistic astrophysical environment, PBHs that form with a non-zero electric charge are expected to quickly neutralize via emission and/or accretion~\cite{Eardley:1975kp}. However, in scenarios Beyond the Standard Model PBHs can hold their charge over cosmological time-scales. We will assume a scenario where there is a new dark $u(1)$ gauge symmetry and a heavy dark electron.  The new Lagrangian terms will be
\begin{align}
    \mathcal{L} 
    \supset
    - \frac{1}{16\pi}F'^{\mu\nu}F'_{\mu\nu}
    +
    \overline{\ell_D} (i\slashed{D} - m_D)\ell_D
    \,,
\end{align}
where $F'_{\mu\nu} = \partial_\mu A'_\nu - \partial_\nu A'_\mu$ is the field strength tensor of the dark photon, $D_\mu = \partial_\mu + i e_D A'_\mu$ is a covariant derivative, $m_D$ and $e_D$ are the mass and dark charge of the dark electron, respectively, and we use units where the dark vacuum permittivity is $\epsilon_0' = 1/4\pi$.  For simplicity, we will take the kinetic mixing term $\varepsilon F_{\mu\nu}F'^{\mu\nu}$ to be negligible and the dark photon to be massless, see the Appendix for a more detailed discussion on this.  In this scenario \cref{eq:Tbh,eq:d2NdEdt} still apply, with $Q$ replaced by the dark charge $Q_D$.  We use the greybody factors provided by \texttt{BlackHawk}~\cite{Arbey:2019mbc,Arbey:2021mbl} up to the maximum value of the charge parameter for which they are available, $Q_D^{*}=0.999$. For larger values we use the prescription described in~\cite{Baker:2025zxm}.  

Assuming some PBHs start with a small dark charge parameter $Q^*_D \ll 1$, they would then lose mass due to Hawking radiation (e.g., via massless photons) but if the dark electron mass is much larger than the PBH temperature, then dark electrons will not be emitted and the dark charge of the PBH will not change.  This leads to the PBH becoming quasi-extremal (i.e., $Q^*_D \sim 1$) shortly after formation.  Once they become quasi-extremal their Hawking radiation is heavily suppressed, both through temperature suppression and through the greybody factor, and they become meta-stable. Eventually, however, the dark electric field near the event horizon becomes strong enough for the (dark) Schwinger effect~\cite{Schwinger:1951nm} to become active, which quickly discharges the PBH. The black hole then explodes like a Schwarzschild black hole. During the last instants of the explosion, PBHs emit high-energy photons, which can be used to constrain their abundance~\cite{Carr:2020gox}, as well as high-energy neutrinos.  For certain values of $e_D$ and $m_D$, a quasi-extremal PBH could discharge at the right time so that its emission around 1\,PeV is heavily suppressed while its emission around 100\,PeV is less suppressed.  In this way this scenario can explain why the burst rate implied by IceCube seems to be smaller than that implied by KM3NeT.

\section{The High-Energy Neutrino Flux from a Quasi-Extremal PBH}
\label{Sec:neutrinos}

For a given initial mass and charge $Q_D^{*i}$, the differential equations outlined below can be solved to find the evolution of mass and charge of a PBH with time.  Given this evolution, \cref{eq:d2NdEdt} can be used to find the spectrum of the Hawking radiation as a function of time.

As well as directly emitting primary neutrinos as Hawking radiation, other emitted particles may produce secondary neutrinos as they hadronise and decay.  Assuming that primary neutrinos are produced in their flavour eigenstates, the spectrum of neutrinos of flavour $\beta$ produced by the PBH is
\begin{align}
    \frac{d^2N_{\text{total}}^{\nu_\beta}}{dtdE}\bigg |_\text{PBH}
    =
    \sum_i
    &
    \int_0^\infty
    \!dE'\,
    \frac{d^2N_p^i}{dtdE'}
    \frac{dN^{i\to \nu_\beta}(E,E')}{dE}
    +
    \notag\\
    &
    +\mathcal{A}^{\nu_\beta\to\nu_\beta}(E)
    \frac{d^2N_p^{\nu_\beta}}{dtdE}\,,
\end{align}
where the first term accounts for the secondary neutrinos, which is obtained by integrating the primary spectrum of particle $i$ against their fragmentation functions $dN^{i\to \nu_\beta}(E,E')/dE$, and where the second term is the fraction of primary neutrinos which does not undergo any fragmentation process, given by the survival factor $\mathcal{A}^{\nu_\beta\to\nu_\beta}$.  We use the public code \texttt{HDMSpectra} \cite{Bauer:2020jay} to obtain the fragmentation functions and survival factors.
 
After production, the neutrinos travel to Earth in their mass eigenstates before being detected in their flavour eigenstates.  The final neutrino spectrum of flavour $\alpha$ at Earth is~\cite{Capanema:2021hnm}
\begin{align}
    \frac{d^2N_\text{total}^{\nu_\alpha}}{dtdE}\bigg |_\text{Earth}
    &=
    \sum\limits_{\beta}\,
    \sum\limits_{i=1}^3
    |U_{\alpha i}|^2|U_{\beta i}|^2
    \frac{d^2N_{\text{total}}^{\nu_\beta}}{dtdE}\bigg |_\text{PBH}\,,
    \label{eq:neutrinospectrum}
\end{align}
where $\beta\in\{e,\mu,\tau\}$ and $U$ is the PMNS matrix.   

\begin{figure}
    \centering
    \includegraphics[width=0.5\textwidth]{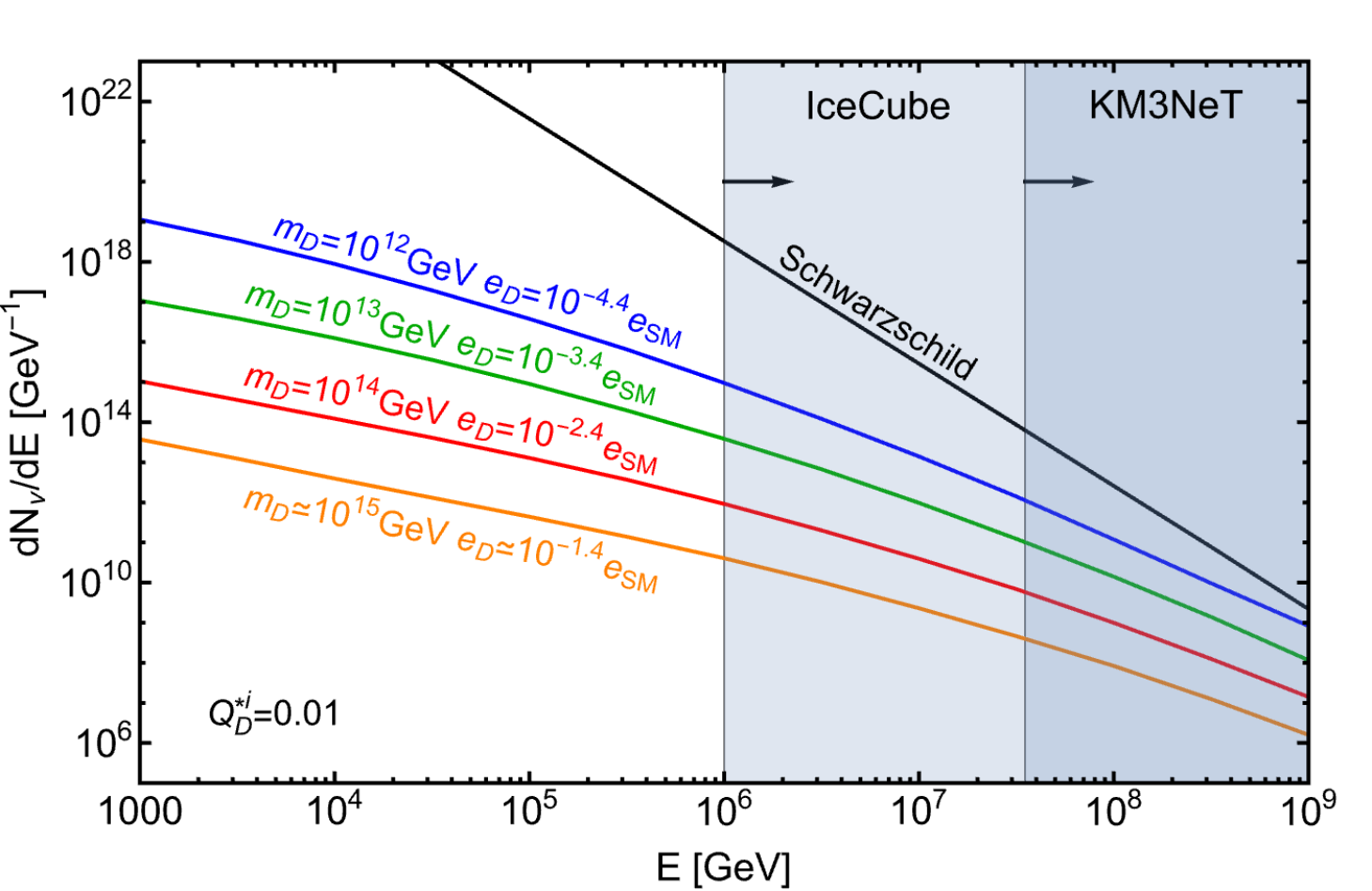}
    \caption{Time integrated neutrino spectra during the explosion of Schwarzschild (black) and Reissner–Nordstr\"om (colour) black holes for different dark electron masses and couplings. We take $Q_D^{*i} = 0.01$ and $M_\text{PBH}^i= 5.6\times 10^{14}$\,g, $3.5\times 10^{7}$\,g, $3.7\times 10^{6}$\,g, $3.9\times 10^{5}$\,g and $7.2\times 10^{4}$\,g for the black, blue, green, red and orange curves, respectively. We integrate over the last 0.09s, 0.62s, 0.48s, 0.23s and 0.15s, respectively, according to the prescription described in the text. The blue bands indicate the IceCube and KM3NeT neutrino energy ranges.}
    \label{fig:NnuIntegrated}
\end{figure}

The number of neutrinos above an energy $E_{\rm min}$ emitted by a single PBH over a time interval $\Delta t = \tau_{\rm max}-\tau_{\rm min}$, evaluated at the distance of the Earth from the PBH, is
\begin{align}
    N^{\nu_\alpha}
    =
    \int_{E_{\text{min}}}^{\infty}
    \!dE\,
    \int_{\tau_{\text{min}}}^{\tau_{\text{max}}}
    \!d\tau
    \frac{d^2 N^{\nu_\alpha}_\text{total}}{dtdE}\bigg|_\text{Earth}.
    \label{eq:EBHSignal}
\end{align}
We take $E_{\rm min}=1$\,PeV for IceCube and $E_{\rm min}=35$\,PeV for KM3NeT, which correspond to the typical energy scale of the IceCube events and to the lower 90$\%$ confidence interval for the muon energy triggering the KM3-230213A event (the muon energy serves as a lower limit on the incoming neutrino energy~\cite{KM3NeT:2025npi}), respectively.
To capture the high-energy neutrinos we integrate from the time the neutrino spectrum in \cref{eq:neutrinospectrum} at $E=10^6$\,GeV is equal to 10\,$\text{GeV}^{-1} \text{s}^{-1}$ and until the time the PBH mass reaches $1$g.

In \cref{fig:NnuIntegrated} we show the time-integrated total neutrino spectrum for a Schwarzschild black hole and for quasi-extremal PBHs for different dark electron masses and dark charges (which we show for the smallest dark charge for which our equations are valid).  We choose the initial mass $M_\text{PBH}^i$ so that their lifetimes are the age of the universe $t_{\rm U} = 1.38 \times 10^{10}$\,yr.  We can see that the RN PBHs have a suppressed emission compared to the Schwarzschild case.  There is a larger suppression in the energy range where IceCube has reported several detections compared to the energy range of the KM3NeT observation.  For a dark electron mass of $10^{12}$\,GeV the neutrino flux at $\sim 10^9$\,GeV is similar for quasi-extremal and Schwarzschild black holes as the PBH discharges when $T_\text{PBH} \sim 10^9$\,GeV.  As the dark electron mass increases, this happens at higher energies as the discharge mass ($M_{d} \equiv e_D M_\text{Pl}^3/\pi m_D^2$) decreases.

\section{Inferred Burst Rates by KM3NeT and IceCube}
\label{Sec:burst-rate}

Going beyond a single PBH, we now study whether a population of dark charged PBHs can simultaneously explain the KM3NeT and IceCube observations, while also satisfying the indirect EGRB limit.  We first assume a population of PBHs with the same initial mass, fixed so their lifetime is $t_{\rm U}$, and dark charge parameter, $Q_D^{*i} = 0.01$.

Neglecting the contribution from extragalactic PBHs \cite{Klipfel:2025jql}, we compute the expected high-energy neutrino flux at Earth from a population of galactic PBHs. We assume the PBHs are distributed like dark matter, which we take to follow a modified Navarro-Frenk-White profile~\cite{Navarro:1995iw},
\begin{align}
    \rho_{\rm DM}(r,z) = \frac{\rho_0}{L(r,z) (1+L(r,z))^2} e^{-\left(\frac{r_0 L(r,z)}{r_{\rm vir}}\right)^2},
\end{align}
where we use cylindrical coordinates in the Galactocentric frame, $L(r,z) \equiv [(r/r_0)^2+(z/[q \,r_0])^2]^{1/2}$ and the best-fit parameters are $\rho_0=0.0196 M_{\odot}$\,pc$^{-3}$, $r_0=15.5$\,kpc, $r_{\rm vir}=287$\,kpc and $q=1.22$~\cite{Binney_2017,Posti_2019}. The burst rate at a position $\mathbf{x}$ will then be~\cite{Klipfel:2025jql}
\begin{align}
    \dot n_\text{PBH}(\mathbf{x})
    = 
    \frac{\rho_\text{DM}(\mathbf{x})}{\rho_\text{DM}(\mathbf{x}_\odot)} 
    \dot n_\text{PBH}(\mathbf{x}_\odot)
    \,,
\end{align}
where $|\mathbf{x}_\odot|=R_\odot=8.3$\,kpc, $\rho_\text{DM}(\mathbf{x}_\odot)$ is the local dark matter energy density and $\dot n_\text{PBH}(\mathbf{x}_\odot)$ is the PBH burst rate in the vicinity of Earth.

Assuming that KM3NeT and IceCube uniformly sample the whole sky over their observation periods, so the neutrino flux can be taken to be isotropic, the expected neutrino flux (in units of s$^{-1}$cm$^{-2}$sr$^{-1}$) at energies above $E_{\rm min}$ at Earth is given by 
\begin{align}
    \phi^{\rm iso}_\nu (E_{\rm min}) 
    &=
    \frac{1}{4 \pi}\int_V
    \frac{\sum_\alpha N_{\nu_\alpha}}{4\pi |\mathbf{x} - \mathbf{x}_\odot|^2}
    \dot n_\text{PBH}(\mathbf{x})
    dV \,,
\end{align}
where $V$ is a sphere of radius $r_\text{vir}$ centred on the Milky Way, $\alpha\in\{e,\mu,\tau\}$ and we include both neutrinos and anti-neutrinos (as KM3NeT and IceCube are sensitive to electron, muon and tau neutrinos and anti-neutrinos).

\begin{figure}
    \centering
    \includegraphics[width=0.5\textwidth]{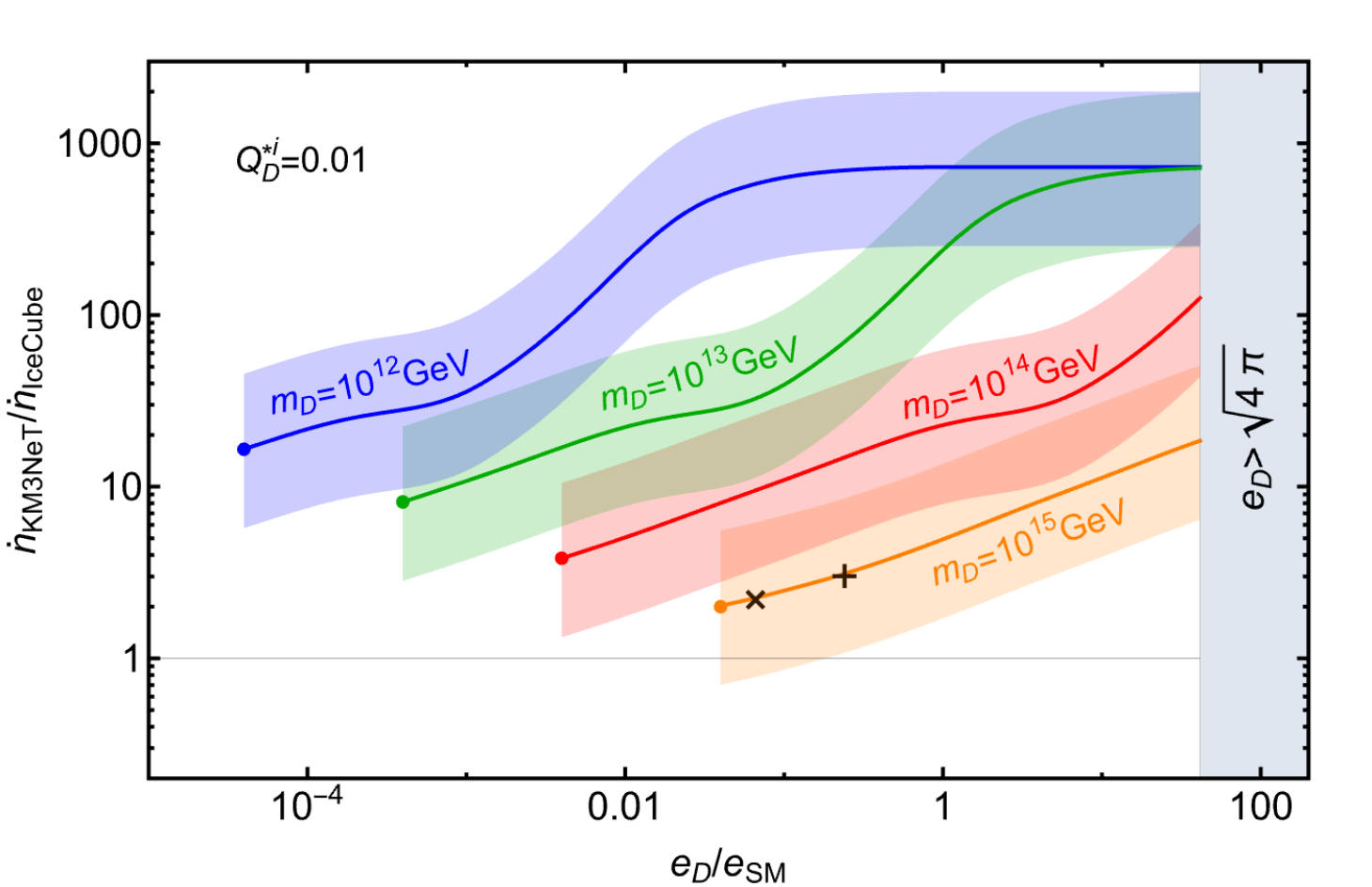}
    \caption{Ratio of the burst rate inferred by the KM3NeT flux over that inferred by the IceCube flux for several dark electron masses, as a function of the dark coupling. The coloured bands correspond to the $1\sigma$ error. The coloured dot indicates the point beyond which our evolution equations are not applicable, while perturbativity requires $e_D<\sqrt{4\pi}$.}
    \label{fig:BurstRatio}
\end{figure}

We can then use the neutrino fluxes reported by KM3NeT~\cite{KM3NeT:2025npi} and IceCube~\cite{Naab:2023xcz} to fit $\dot n_\text{PBH}(\mathbf{x}_\odot)$, using the corresponding choice for $E_{\rm min}$. 
Assuming a population of Schwarzschild PBHs, this leads to local burst rates of $\dot{n}_{\rm PBH} = 1.00^{+1.75}_{-0.64} \times 10^6 \,\text{pc}^{-3}\text{yr}^{-1}$ for KM3NeT and $\dot{n}_{\rm PBH} = 1.38^{+0.19}_{-0.19} \times 10^3 \,\text{pc}^{-3}\text{yr}^{-1}$ for IceCube, which differ by three orders of magnitude. 
In \cref{fig:BurstRatio} we show our results as the ratio of burst rates inferred by KM3NeT and IceCube as a function of the dark coupling.
We see that the ratio is unchanged for large dark couplings and $m_D \lesssim 10^{13}$\,g. For these parameters the PBHs are Schwarzschild-like when emitting in the 1 -- 100\,PeV range.  
However, as we move to lower values of $e_D$, it is possible for the burst rates to agree within $1\sigma$ for $m_D=10^{15}$\,GeV. If the dark coupling gets too small then our evolution equations do not apply, see below.  Although our code does not converge in a reasonable time for $m_D \gtrsim10^{15}$\,GeV, we note that the trend seen in \cref{fig:BurstRatio} seems to indicate that there are more points in the dark sector parameter space that can yield a consistent burst rate ratio for larger dark electron masses. Overall, we see that this scenario can achieve a very significant relaxation of the tension between KM3NeT and IceCube.

\begin{figure}
    \centering
    \includegraphics[width=0.5\textwidth]{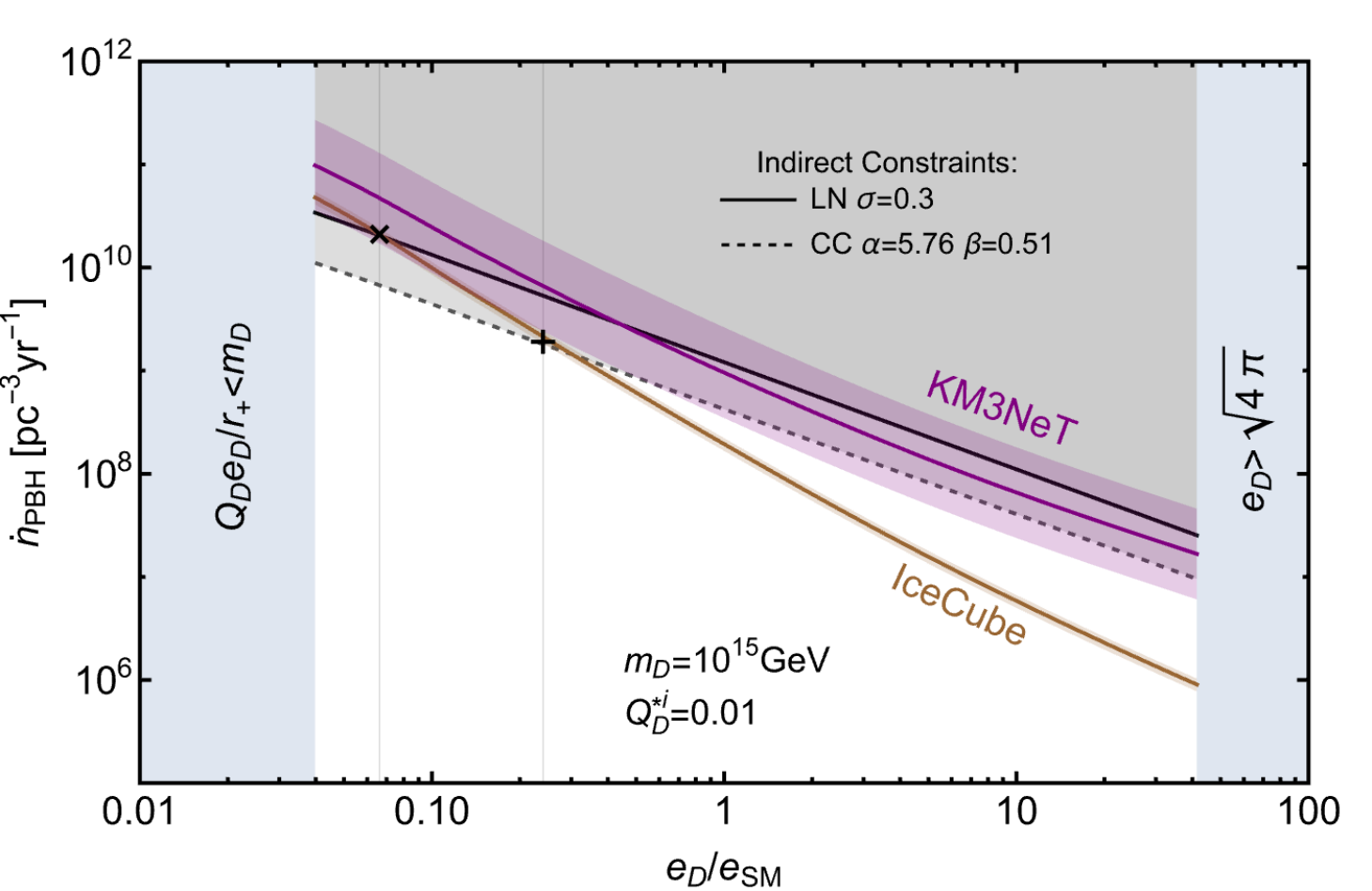}
    \caption{Burst rates inferred by KM3NeT (purple) and IceCube (brown) with their $1\sigma$ error bands for $m_D=10^{15}$GeV. The black lines indicate the maximum allowed burst rate by indirect constraints assuming a log-normal (LN) mass distribution with $\sigma=0.3$ and a generalized critical collapse (CC) distribution with $\alpha=5.76$ and $\beta=0.51$~\cite{Gow:2020cou}.  The $\times$ and $+$ indicate the same parameter points as in \cref{fig:BurstRatio}.}
    \label{fig:BurstRates}
\end{figure}

We now consider physically motivated PBH mass distributions. We first note that the KM3NeT and IceCube burst rates inferred for Schwarzschild PBHs are significantly larger than the local burst rates allowed by indirect constraints from the EGRB, $\dot{n}_{\rm PBH} \lesssim 0.01-0.1 \,\text{pc}^{-3}\text{yr}^{-1}$~\cite{Boluna:2023jlo}. Furthermore, the burst rate inferred from the KM3NeT observation is not consistent with the upper limit derived by HAWC, $\dot{n}_{\rm PBH} \lesssim 3400 \,\text{pc}^{-3}\text{yr}^{-1}$~\cite{HAWC:2019wla}.  In \cref{fig:BurstRates} we show the burst rates for quasi-extremal PBHs along with the indirect EGRB and cosmic microwave background constraints~\cite{Carr:2009jm} (following Refs.~\cite{Carr:2017jsz,Boluna:2023jlo,Baker:2025zxm}) for a log-normal and a generalized critical collapse mass distribution~\cite{Gow:2020cou}, where we allow the peak mass to float.  We see that as the dark coupling reduces, the inferred burst rates both increase, but the IceCube rate increases faster.  The indirect constraints also weaken, and at $e_D \sim 0.07 e_\text{SM}$ for the log-normal distribution and $e_D \sim 0.24 e_\text{SM}$ for generalized critical collapse, indicated by $\times$ and $+$ respectively, all three are consistent to within $\sim1\sigma$.  While the inferred burst rate is quite large, around $10^{10}\,\text{pc}^{-1}\text{yr}^{-1}$, each burst produces much less low energy Hawking radiation than in the Schwarzschild case so the indirect constraints are still satisfied.  The direct constraints are also likely to be satisfied, see below.  Interestingly, for the log-normal mass distribution with a width of $\sigma = 0.3$ we find the best agreement when $f_\text{PBH} \sim 1$ (with a peak mass of $M^* \simeq 3.2 \times 10^{5}$g), indicating that this population of PBHs could constitute the entirety of dark matter in the universe (although we find good agreement for $5\times10^{-7} \lesssim f_\text{PBH} \leq 1$ corresponding to peak masses between $M^* \simeq 1.4 \times 10^{4}$g and $M^* \simeq 3.2 \times 10^{5}$g, and for the critical collapse distribution shown the best agreement is for $f_\text{PBH} \sim 0.01$).  Finally, as discussed in Ref.~\cite{Baker:2025zxm}, we do not expect a significant change in the results when considering a range of initial dark charge parameters.

\section{Conclusion}
\label{Sec:conclusions}

If there is a new dark sector and PBHs formed with a small dark charge, then light PBHs could become quasi-extremal before building up a strong enough dark electric field to rapidly discharge via dark Schwinger emission.  In this letter we have shown that the neutrino emission in the PeV range can then be suppressed relative to that in the 100 PeV range.  This could explain the high energy neutrino fluxes seen in IceCube and KM3NeT, while remaining consistent with direct and indirect constraints.  Intriguingly, a log-normal mass distribution with $\sigma = 0.3$ is consistent with these PBHs constituting the dark matter in the universe.

\section*{Acknowledgements}

This work was supported by the University of Massachusetts, Amherst.  We would also like to acknowledge the Amherst Center for Fundamental Interactions.

\appendix

\section{The Mass and Charge Evolution of a Dark Charged PBH}

We now describe the time evolution of a charged black hole in the limits where the (reduced) Compton wavelength of the dark electron is much smaller than the black hole radius, $1/m_D \ll M_\text{PBH}/M_{\rm Pl}^{2}$, and where the potential energy gained by a dark electron that is repelled from the black hole horizon to infinity is much greater than its rest mass, $Q_D e_D/r_+ \gg m_D$.\footnote{Note that this limit is slightly stronger than the weak gravity conjecture, $e_D \gtrsim  m_D/M_\text{Pl}$.}  In this case the black hole mass evolution is given by~\cite{Hiscock:1990ex}
\begin{align}
\label{eq:m-evolution}
    \frac{dM_{\rm PBH}}{dt} &= -\frac{\alpha(M_{\rm PBH}, Q_D)}{M^{2}_{\rm PBH}}+\frac{Q_D}{r_{+}}\frac{dQ_D}{dt}\,,
\end{align}
where
\begin{align}
    r_{+}=\frac{M_{\rm PBH}}{M_{\rm Pl}^{2}}\left(1 + \sqrt{1 - 
    (Q_D^\ast)^{2}
    }\right)
\end{align}
is the outer horizon radius of an RN black hole.  The Page function $\alpha(M_{\rm PBH}, Q_D)$ is
\begin{align}
    \alpha(M_{\rm PBH}, Q_D) &= M_{\rm PBH}^{2} \sum_{i} \int_{0}^{\infty} E \frac{d^{2}N_p^{i}}{dEdt} \, dE 
    \,,
\end{align}
where the sum is over all particles present in nature.


To find $d Q_D/dt$ we begin with the leading order Schwinger equation for the rate of dark electron positron pair production per unit four volume~\cite{PhysRevLett.33.558,Hiscock:1990ex,Cohen:2008wz,Brown:2024ajk},
\begin{align}
    \Gamma(r)
    &=
    \frac{e_D^2}{4\pi^3}
    \frac{Q_D^2}{r^4}
    \exp{\left(
        \frac{-\pi m_D^2 r^2}{e_D Q_D}
    \right)}
    \,.
\end{align}
%
Particles with an opposite charge to the PBH will be attracted back into the PBH and do not contribute to mass and charge loss. To find the number of dark electrons emitted over a short time, $\Delta N_D$, we integrate over the space outside the outer-horizon and over the time interval $\Delta t$,
\begin{align}
    \Delta N_D
    &=
    \int_{r\geq r_+, \Delta t}
    d^4x \sqrt{|g|}
    \Gamma(r)
    \\
    &=
    \frac{e_D^2 Q_D^2}{\pi^2} \Delta t
    \int^\infty_{r_+} dr
    \frac{1}{r^2}
    \exp{
    \left(\frac{-\pi m_D^2r^2}{e_D Q_D}\right)}
    \,,
\end{align}
where $g$ is the determinant of the RN metric, 
\begin{align}
    g_{\mu\nu} &= \text{diag}(-f(r), 1/f(r), r^2, r^2\sin^2(\theta))
\end{align}
in the co-ordinates $(t,r,\theta,\phi)$, and where
\begin{align}
    f(r) = 1-\frac{2M_\text{PBH}}{r M_\text{Pl}^2} + \frac{Q_D^2}{r^2 M_\text{Pl}^2}
    \,.
\end{align}
Evaluating the integral 
and taking the limits $\Delta t\to dt$ and $\Delta N_D \to d N_D$ yields \cite{CamposDelgado:2023rti}
\begin{align}
    \frac{dQ_D}{dt}
    =
    -e_D
    \frac{dN_D}{dt}
    =
    -\frac{e_D^3 Q_D^2}{\pi^2}
    \left[
        \frac{e^{-\frac{r_+^2}{a}}}{r_+}
        -
        \sqrt{\frac{\pi}{a}}
        \text{Erfc}\left(\frac{r_+}{\sqrt{a}}\right)
    \right]
    \,,
\end{align}
where $a = e_D Q_D/\pi m_D^2$ and Erfc is the complementary error function.


\section{Associated Gamma Ray Signal}

A neutrino event from an exploding PBH is expected to be associated with a burst of high energy photons which could be observed by existing gamma ray telescopes, such as HAWC or LHAASO. The location of the KM3-230213A event was such that the PBH was within HAWC's field of view at the time of the final explosion and within LHAASO's field of view seven to fourteen hours before the final burst \cite{Airoldi:2025opo}. Unfortunately, the HAWC telescope was not operational at the time of the event. However, if the KM3-230213A event was to be explained by a standard Schwarzschild PBH, it should have produced many detectable gamma rays when it was in LHAASO's field of view seven to fourteen hours prior to the explosion \cite{Airoldi:2025opo}, which were not observed. In our scenario, the final burst of a quasi-extremal PBH happens much faster than for a Schwarzschild black hole. We estimate that the emission of gamma-rays within the range of sensitivity of LHAASO will only take place within the final hundreds of seconds before the final explosion. The final burst of a quasi-extremal PBH is thus not expected to be observable hours before the final explosion. Quasi-extremal black holes thus resolve the tension between Schwarzschild black holes and the non-observation of gamma-rays by LHAASO. 

In the future, a concurrent gamma-ray signal would be expected if the event occurs within HAWC's or LHAASO's field of view. However, their sensitivity to extremely high-energy photons above $\sim 500\,$TeV and $\sim 1000\,$TeV, respectively, remains unclear since above these energies the Cherenkov detectors become saturated. While a higher energy event would still have an effect on the telescope, its energy and direction could not be confidently measured. Furthermore, the ultra-high energy cosmic ray flux is a large background and at these high energies HAWC's ability to distinguish cosmic rays and gamma rays is unclear. The Pierre Auger Observatory as well as future telescopes, such as SWGO and CTA, may also be sensitive.  However, a dedicated analysis focussing on associated high-energy gamma rays is beyond the scope of this letter and left to future work.

\section{Non-Zero Kinetic Mixing}

While for simplicity we have assumed that the kinetic mixing, $\varepsilon$, between the photon and dark photon is negligible, we now consider the impact of a non-zero $\varepsilon$. Here we consider the case of a massless dark photon.

Prior to mixing, the relevant terms are~\cite{Fabbrichesi_2021}
\begin{align}
    \mathcal{L}
    \supset
    &-\frac{1}{16\pi}F_a^{\mu\nu}F_{a\mu\nu}
    -\frac{1}{16\pi}F_b^{\mu\nu}F_{b\mu\nu}
    -\frac{\varepsilon}{8\pi}F_a^{\mu\nu}F_{b\mu\nu}\notag\\
    &+e_D J'_{\mu} A_a^\mu
    +e J_{\mu} A_b^\mu
    \,,
\end{align}
where $F_{a,b}^{\mu\nu}$ are the field strength tensors for the two Abelian gauge fields $A_{a,b}^\mu$, $\varepsilon$ is the kinetic mixing parameter, and $J^\mu$ and $J'_\mu$ are the SM and dark fermion currents coupled to the gauge bosons, respectively.  We can diagonalise the gauge kinetic terms by performing the field redefinition
\begin{align}
    \begin{pmatrix}
        A_a^\mu \\ A_b^\mu
    \end{pmatrix}
    =
    \begin{pmatrix}
        \frac{1}{\sqrt{1-\varepsilon^2}} & 0\\
        -\frac{\varepsilon}{\sqrt{1-\varepsilon^2}} & 1
    \end{pmatrix}
    \begin{pmatrix}
         \cos \theta & -\sin \theta\\
         \sin \theta & \cos \theta 
    \end{pmatrix}
    \begin{pmatrix}
        A'^\mu \\ A^\mu
    \end{pmatrix}
    \,,
\end{align}
where $A'^\mu$ is the dark photon and $A^\mu$ the SM photon.  Different choices of $\theta$ are physically equivalent for a massless dark photon but result in different couplings between the gauge fields and the fermion currents.  

In the basis with $\sin\theta=\varepsilon$ the SM electron couples to the SM photon and has SM charge $e/\sqrt{1-\varepsilon^2}$ while the dark electron couples to both the dark photon with dark charge $e_D$ and to the SM photon with milli-charge $-\varepsilon e_D/\sqrt{1-\varepsilon^2}$.  We then imagine that a PBH formed while accreting $N'$ more dark electrons than dark positrons so the charges of the PBH are
\begin{align}
    \left(Q_D,\,Q\right) = N'\left(e_D\,,-\frac{\varepsilon e_D}{\sqrt{1-\varepsilon^2}}\right)
    \,.
\end{align}
The PBH can then efficiently discharge the SM electric charge shortly after formation by emitting $N=N'\varepsilon e_D/ e$ more positrons than electrons, so that the PBH becomes SM neutral.  Since the SM electrons are not milli-charged in this basis, the final charges of the PBH are
\begin{align}
    \left(Q_D,\,Q\right) = N'(e_D,0)
    \,.
\end{align}

In the basis with $\sin\theta=0$ the SM electron couples to the SM photon with SM charge $e$ and to the dark photon with milli-charge $-\varepsilon e/\sqrt{1-\varepsilon^2}$ while the dark electron only couples to the dark photon with dark charge $e_D/\sqrt{1-\varepsilon^2}$.  We again imagine that a PBH forms accreting $N'$ more dark electrons than dark positrons so the PBH charges are
\begin{align}
    \left(Q_D,\,Q\right) = N'\left(\frac{e_D}{\sqrt{1-\varepsilon^2}}\,,0\right)
    \,.
\end{align}
Since SM electrons now have a dark milli-charge the PBH preferentially radiates SM positrons over SM electrons. The PBH then builds up a negative SM charge,
\begin{align}
    \left(Q_D,\,Q\right) = \left(\frac{N'e_D-N\varepsilon e}{\sqrt{1-\varepsilon^2}},Ne\right)
    \,,
\end{align}
where $N$ is the excess number of SM positrons emitted over SM electrons. The potential energy $U$ on a SM positron at a distance $r$ from the centre of the PBH due to the PBH's SM and dark charges is
\begin{align}
    U
    &=
    \frac{\varepsilon e}{\sqrt{1-\varepsilon^2}}
    \frac{Q_D}{r}
    -
    e\frac{Q}{r}
    \\
    &=
    \frac{1}{r}
    \left(
    \frac{\varepsilon e}{1-\varepsilon^2}
    \left(
    N' e_D
    -
    N\varepsilon e
    \right)
    -
    Ne^2
    \right)
    \\
    &=
    \frac{e}{(1-\varepsilon^2) r}
    \left(
    N' \varepsilon e_D
    -
    Ne
    \right)
    \,,
\end{align}
so after $N=N'\varepsilon e_D/e$ excess positrons are emitted the potential is constant.  The PBH then no longer preferentially emits SM positrons or electrons.

We see that in both bases the physics is the same: a PBH, formed by accreting more dark electrons than dark positrons, quickly preferentially emits $N = N'\varepsilon e_D/e$ SM positrons before reaching a state where it has a dark charge and does not preferentially emit SM electrons or positrons.  The result is a dark charged PBH that does not `leak' its charge by SM emission, irrespective of the size of the kinetic mixing parameter $\varepsilon$.  Interestingly, the charge of the PBH changes between the two bases (although $Q_D^2 + Q^2$ is the same).

In the case of a massive dark photon a more detailed analysis needs to be performed, taking into account the fact that dark electromagnetism then becomes a short range force.


\bibliography{biblio}

\end{document}